# Relocalization of Uranium $5f$ Electrons in Antiferromagnetic Heavy Fermion Superconductor UPd$_2$Al$_3$: Insights from Angle-Resolved Photoemission Spectroscopy

Jiao-Jiao Song,[1] Chen Zhang,[1] Qi-Yi Wu,[1] Yin-Zou Zhao,[1] Ján Rusz,[2] John J. Joyce,[3] Kevin S. Graham,[3] Peter S. Riseborough,[4] Clifford G. Olson,[5] Hao Liu,[1] Bo Chen,[1] Ya-Hua Yuan,[1] Yu-Xia Duan,[1] Paul H. Tobash,[3] Eric D. Bauer,[3] Peter M. Oppeneer,[2] Tomasz Durakiewicz,[6] and Jian-Qiao Meng[1, *]

[1]*School of Physics, Central South University, Changsha 410083, Hunan, China*
[2]*Department of Physics and Astronomy, Uppsala University, Box 516, S-75120 Uppsala, Sweden*
[3]*Los Alamos National Laboratory, Los Alamos, New Mexico 87545, USA*
[4]*Temple University, Philadelphia, Pennsylvania 19122, USA*
[5]*Ames Laboratory, Iowa State University, Ames, Iowa 50011, USA*
[6]*Idaho National Laboratory, Idaho Falls, Idaho 83415 USA*
(Dated: Tuesday 30th April, 2024)

We investigate the antiferromagnetic heavy fermion superconductor UPd$_2$Al$_3$, employing angle-resolved photoemission spectroscopy to unravel the complex electronic structure of its U $5f$ electrons. We observe unexpected characteristics that challenge the conventional temperature-dependent behavior of heavy fermion systems, revealing unexpected characteristics. At temperatures above the anticipated coherence temperature ($T^*$), we observe itinerant U $5f$ electrons at temperatures higher than previously postulated. Additionally, a previously unidentified dispersionless band emerges around 600 meV below the Fermi energy, potentially linked to spin-orbit splitting within the U $5f$ states. Hybridization between the $5f$ electrons and conduction band was observed with an energy dispersion of 10 meV at low temperatures, suggesting that U $5f$ electrons near and at the Fermi surface have an itinerant nature. Temperature-dependent $5d$-$5f$ resonance spectra reveal that the $5f$ electron spectrum weight increases with lowering temperature and begins to decrease at temperatures significantly higher than the Néel temperature ($T_N$). We further show that the competition between the Kondo effect and Ruderman-Kittel-Kasuya-Yosida (RKKY) interactions may be responsible for the relocalization of $5f$ electrons, making relocalization a precursor to the establishment of magnetic order at lower temperatures. Our experiments also provide evidence that $5f$ electrons with the same orbital are involved in both the Kondo effect and RKKY interactions, suggesting that the two coexist at lower temperatures.

## I. INTRODUCTION

Heavy Fermion (HF) compounds have garnered significant attention for their distinctive physical properties and abundance of exotic states, including quantum criticality, unconventional superconductivity, and coexisting magnetic orderings [1, 2].The ground state of HF systems is determined by the competition between Kondo interaction and RKKY interaction [3]. In a traditional view, the nature of $4f$ and $5f$ electrons in HF compounds is believed to depend on temperature: above the coherence temperature $T^*$, $f$ electrons are considered completely localized, and upon cooling below $T^*$, localized $f$ electrons start to hybridize with conduction electrons, leading to their delocalization, and finally, at low temperatures, to coexistence of itinerant and localized $f$ electrons [4–7].

Recent investigations, particularly employing angle-resolved photoemission spectroscopy (ARPES), have challenged these traditional views. In certain Ce- and Yb-based HF systems, $4f$ bands have been observed at temperatures well above $T^*$, implying the presence of itinerant $f$ electrons [8–16]. Ultrafast optical spectroscopy experiments revealed hybridization fluctuations in these materials before the establishment of heavy elec-tron coherence [17, 18], resulting in the opening of a direct energy gap at the coherence temperature [17–19]. Other studies have suggested that hybridization can be suppressed at lower temperatures, causing relocalization of Ce $4f$ electrons [14–16]. Notably, the Ce $4f$ band observed near the Fermi energy ($E_F$) by ARPES represents the small tails of the Kondo resonance peak located above $E_F$. In the case of U-based HF compounds, the significant number of $5f$ electrons at $E_F$ [20–28] provides an ideal platform to explore the nature of $f$ electrons and the intricate interplay between the Kondo effect and RKKY interaction.

UPd$_2$Al$_3$ is a prototypical HF compound with a large specific heat coefficient ($\gamma$) of 210 mJ/mol·K$^2$ [29]. UPd$_2$Al$_3$ stands out as the first reported HF material in which superconductivity and antiferromagnetism coexist. It undergoes an antiferromagnetic (AFM) phase transition at the $T_N$ of 14 K, followed by a transition to an unconventional superconducting state at a lower critical temperature ($T_c$) of 2 K [29–31]. ARPES and optical conductivity measurements helped identify the onset of a hybridization gap at $T^* \sim 65$ K [20] or $\sim 50$ K [32], respectively. Quasiparticle scattering spectroscopy observed the persistence of hybridization gap up to around 28 K [33]. Nevertheless, the nature of the U 5$f$ electrons in UPd$_2$Al$_3$ remains a subject of debate. De Haas-van Alphen experiments at low temperatures are consistent with an itinerant 5$f$-electron model [7, 34], and photoemission studies also suggest the itinerant nature of the 5$f$ state at low temperatures [20–23]. However, incoherent peaks observed in resonance photoemission have been interpreted as a sign of partially localized U 5$f$ electrons [23]. Temperature-dependent magnetic susceptibility and heat capacity measurements have suggested that the 5$f$ electrons tend to be almost completely localized, pointing to the presence of crystalline-electric-field (CEF) excitations [35]. The CEF approach has led to proposals of localized 5$f^2$ [29, 35, 36] or 5$f^3$ [37, 38] configurations. To reconcile these seemingly contradictory physical properties, researchers have proposed the concept of a dual nature of U 5$f$ states, combining an itinerant $f^1$ component with a localized $f^2$ component within the U 5$f$ states [33, 39, 40].

## II. EXPERIMENTAL DETAILS

The work presented here was designed to provide the needed experimental evidence to test the hypothesis of 5$f$ duality in UPd$_2$Al$_3$, and reveal the nature of interactions leading to the localization - delocalization balance. High-resolution ARPES measurements were performed at the Synchrotron Radiation Center, Stoughton, WI, using an SES 4000 hemispherical electron energy analyzer and a plane grating monochromator (PGM) beamline. All measurements were approximately conducted along the Γ'-K' direction. U 5$f$ electrons properties were investigated by comparing off- and on-resonance spectra ($h\nu = 102$ and 108 eV, respectively) measured at a low temperature of 20 K with a total energy resolution of ∼ 22 meV. To reveal the itinerant-to-localized transition of U 5$f$-electrons, temperature-dependent on-resonant ARPES measurements were performed. For all these measurements, the angular resolution was 0.20°. High-quality single crystals of UPd$_2$Al$_3$ were grown by the Czochralski method.

## III. RESULTS AND DISCUSSIONS

To examine the properties of U 5$f$ electrons, U 5$d$ → 5$f$ on-resonant ARPES measurements were conducted using 108 eV photons, enhancing the photoconduction matrix element of U 5$f$ electrons. Figures 1(a) and 1(b) depict the broad-range on-resonant valence spectra and their corresponding energy distribution curves (EDCs) at 20 K. Notably, for photons around 108 eV, the contribution of the U 5$f$ state dominates, while the contributions of Pd 4$d$, Al 3$s$, Al 3$p$, and U 6$d$ states sequentially decrease, being one to two orders of magnitude smaller

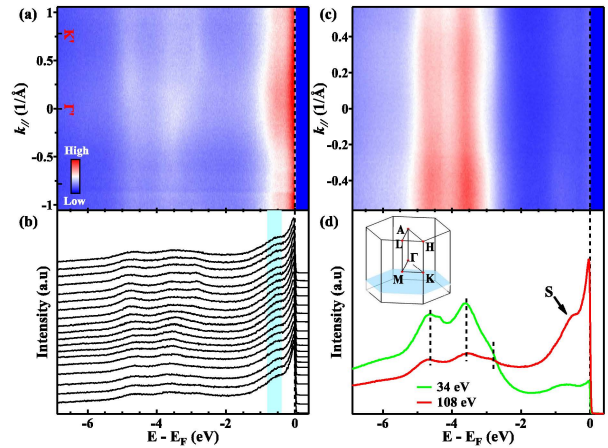

FIG. 1. (color online) Electronic structure of UPd$_2$Al$_3$ measured at 20 K. (a) Raw data of a 2D image showing the broad range on-resonance band structure at 108 eV. (b) EDCs corresponding to (a). (c) Photoemission image taken with 34 eV photons. (d) The red and green curves are the angle-integrated EDCs of (a) and (c), respectively. Inset: Brillouin zone of UPd$_2$Al$_3$ in the paramagnetic phase with high-symmetry momentum points marked, and the ARPES cut plane of 108 eV photons is highlighted in light blue.

[41]. As previously reported, an intense heavy quasiparticle band was observed in the vicinity of $E_F$, signifying the itinerant nature of a significant population of 5$f$ electrons [21–23]. The angle-integrated spectra in red in Fig. 1(d) also exhibit this feature. In contrast to Ce-based HF compounds [42], UPd$_2$Al$_3$ exhibits a significantly higher density of $f$-electron states at $E_F$, indicative of the more itinerant nature of U 5$f$ electrons.

Additionally, we detected a dispersionless feature located at approximately 600 meV below $E_F$, corresponding to the broad satellite peak (S) marked with a cyan shadow in Fig. 1(b) and a black arrow in Fig. 1(d). Satellite peaks with similar energy positions have also been observed in other measurements. The interpretation of these satellite peaks includes their origin from electron correlation effects [21], the lower Hubbard band [22], an unscreened U 5$f^2$-dominant final state [23], or a completely localized U $f^2$ subsystem [43], among other possibilities. X-ray bremsstrahlung isochromat spectroscopy measurements of UPd$_2$Al$_3$ have reported a spin-orbit splitting induced structure above the $E_F$ with an energy of ∼ 0.7 eV [44]. A similar feature, approximately 550 meV below $E_F$, was observed in another U-based HF material, UAsSe [45], which was believed to have some admixture of 5$f$ electrons. According to previous theoretical and experimental results in U-based HFs [45–48], the satellite peak is not a single peak, but rather a multiplet. Upon careful observation of Fig. 1(a), the satellite peak's intensity is strongest near the Γ' point, but notably weaker near the K' point, signifying the existence of large Γ'-centered electron-like bands [34, 48]. Hence, our ARPES data furnishes evidence supporting the interpretation of these satellites as $f^2$ final states multiplet hybridizing with non-$f$ bands [46].

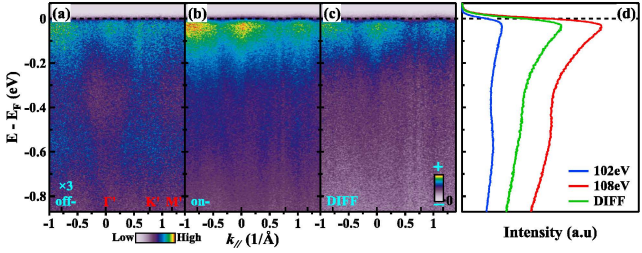

FIG. 2. (color online). (a),(b) Off- and on-resonance, respectively, valence band structure of UPd$_2$Al$_3$ at 20 K with 102 eV and 108 eV. (c) The intensity difference between on- and off-resonance valence band structure. (d) The angle-integrated EDCs for data in (a)-(c).

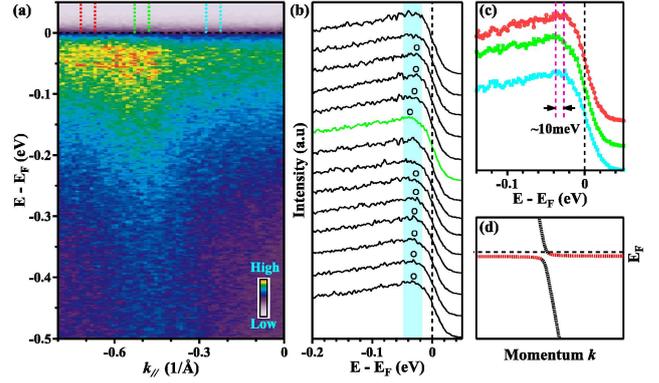

FIG. 3. Formation of the heavy quasiparticle band in UPd$_2$Al$_3$ at 20 K. (a) On-resonant valence band structure of UPd$_2$Al$_3$. (b) EDCs of the spectra shown in (a). The circles indicate the peak position, while the aqua shading indicates itinerant 5$f$ positions. (c) Comparison of EDCs measured at different momentum positions. The red, green, and aqua curves were obtained by integrating the regions between two red, green, and aqua dashed lines, as indicated in (a), respectively. The pink lines serve as a guide for the eyes. (d) Schematic of hybridization.

Furthermore, three less dispersive bands are visible at high binding energies, ranging from -2.5 eV to -5 eV. These features are also observed in the angle-integrated photoemission spectroscopy in red in Fig. 1(d), as indicated by dashed lines. Previous theoretical calculations [21] and soft x-ray measurements [21, 48] have suggested that these bands are primarily derived from the Pd 4$d$ state. For comparison, Fig. 1(c) displays the ARPES image measured with 34 eV photons, resulting in a significant increase in the density of the Pd 4$d$ states. In contrast to the 108 eV photons, the contrast between the photoconduction matrix element of Pd 4$d$ and U 5$f$ states at 34 eV is reversed, with the photoconduction matrix element of Pd 4$d$ being an order of magnitude higher than that of U 5$f$ states [41]. As expected, apart from the observation of the weak flat bands, the Pd 4$d$ states are significantly enhanced. The angle-integrated EDC in green in Fig. 1(d) also exhibits these features.

To gain further insight into the localized or itinerant nature of the 5$f$ electrons, we compared the resonance-enhanced spectra of U 5$f$ weight with spectra dominated by conduction bands. Figures 2(a) and 2(b) display the off- and on-resonance spectra measured at 20 K, respectively. It should be noted that 102 eV is not truly an anti-resonance energy but only a minimum between the two main resonances of 108 eV and 98 eV. In the vicinity of $E_F$, the off-resonance spectrum ($h\nu$ =102 eV) exhibits weaker U 5$f$ orbital character and features very dispersive Al states [Fig. 2(a)]. In contrast, the on-resonance spectrum, resulting from the 5$d$→5$f$ resonance at 108 eV photon energy, displays a significant increase in U 5$f$ spectral weight near $E_F$ [Fig. 2(b)]. The intensities of the heavy quasiparticle bands exhibit strong momentum dependence, evident in both Figs. 2(a) and 2(b), indicating hybridization between the 5$f$ bands and the conduction band.

To elucidate the 5$f$ character, we generated a difference plot between the off- and on-resonance spectra, as shown in Figs. 2(c) and 2(d). The red color near $E_F$ highlights the density of dispersive and hybridized 5$f$ bands contributing to bonding and forming the Fermi surface. This observation contradicts the picture of a localized 5$f^2$ (U$^{4+}$) or 5$f^3$ (U$^{3+}$) configuration [29, 35–38]. Our results provide evidence that the U 5$f$ electrons near the Fermi level are predominantly itinerant at low temperatures.

When hybridization occurs, it leads to the redistribution of $f$ spectral weight and the appearance of dispersive heavy quasiparticle bands near the intersection of the $f$ band and the conduction band [12, 49, 50]. Figure 3 zooms into the vicinity of the Fermi crossing Fig. 2(b), offering a quantitative analysis of hybridization. The hybrid characteristics are evident in Fig. 3(a), where the 5$f$ band at $E_F$ hybridizes with the conduction band, giving rise to a weak dispersive band. This hybridization effect is further illustrated in the EDCs in Fig. 3(b).

For a more detailed look into hybridization, Fig. 3(c) displays angle-integrated EDCs at three selected momentum positions marked with different colors in Fig. 3(a). The selection of these regions is based on the presence of high-intensity dispersion resulting from the hybridization of $f$ and conduction bands crossing $E_F$, as well as low-intensity dispersion areas where no conduction bands cross $E_F$. The peak positions of the EDCs in the nonintersecting regions (red and aqua curves) are nearly identical. While the positions of the two peaks with different dispersion intensities are shifted by a small value of ~10 meV [Fig. 3(c)], similar to that of Ce-based HFs [8, 49] and the directed hybridization gap of URu$_2$Si$_2$ [51]. It should be noted that this value represents the upper limit of the indirected hybridization gap determined by our measurement, which is considerably smaller than the directed hybridization gap reported by Yang *et al.* [20].

To investigate the evolution of U $5f$ electron localization and itinerancy with temperature, we conducted temperature-dependent on-resonance ARPES measurements. Figures 4(a)-(d) display the temperature-dependent on-resonance ARPES results using 108 eV photons. Notably, significant $f$ bands are observed across the entire temperature range. Even at 80 K, which is above the coherence temperature $T^*$, and prior to the opening of the hybridization gap [20], there is already an intense heavy quasiparticle band. Remarkably, at 150 K, well above $T^*$, substantial $f$ bands are still observable (for additional data, see the Supplemental Material [52]). This observation implies that the formation of heavy bands in the U-based HF compound UPd$_2$Al$_3$ begins at temperatures much higher than $T^*$. This finding aligns with recent observations in Ce-based [8–12, 14–16], Yb-based [13], and U-based [25, 26] HFs. It contradicts previous theoretical expectations that heavy electrons should emerge only below the coherence temperature $T^*$, at which hybridization becomes significant [53].

At the lowest temperature measured, 24 K [Fig. 4(a)], we observed an unexpected phenomenon where the intensity of the $f$ band was the weakest among the temperatures depicted. In Fig. 4(e), a comparison of the EDCs measured at 80 K (above $T^*$) and 24 K (well below $T^*$) shows that the main structures are similar. However, in the narrow energy range close to the $E_F$, the EDC intensity at 24 K is slightly lower than that at 80 K, indicating a signature of "relocalization" of the $f$ electrons. This phenomenon will be discussed in more detail later.

In Fig. 4(f), we present the temperature evolution of the integral EDCs over the shown momentum range. Notably, the peak positions of the $5f$ state at $E_F$ and satellite peaks hardly changes with temperature, suggesting that the $5f$ bands do not shift as temperature varies. These findings contrast significantly with those of Fujimori *et al.*, who reported that at high temperatures (100 K), the $5f$ band shifted by approximately 100 meV toward a higher binding energy side and was no longer involved in the formation of the Fermi surface [21]. Our data indicate that the $f$ bands are still present at the $E_F$ at 150 K, well above $T^*$ (for additional data, see the Supplemental Material [52]). Additionally, it can be noted that the EDC intensity shows minimal variation over the presented narrow temperature range. This could be partially attributed to the possible significant contribution of temperature-insensitive, less coordinated surface U atoms to the spectral weight of the U $5f$ state, thereby resulting in weak changes.

Figure 4(g) provides a quantitative representation of the temperature-dependent evolution of the U $5f$ states spectral weight. The spectral weights have been normalized to the data obtained at the highest temperature, 80 K. Notably, the spectral weights of the two $5f$ electron related states exhibit similar temperature-dependent behaviors. It's important to highlight that the spectral weight increases as the temperature decreases initially and then starts to decrease as the temperature drops below approximately $T^*$. This observation indicates that the collective hybridization process begins to reverse, and the Kondo lattice quasiparticles commence relocalizing at low temperatures. Additionally, in Fig. 4(g), we present the temperature dependence of the spectral weight at higher binding energies, integrated over the range of [-1.8 eV, -1.3 eV], as indicated by green diamonds. This approach excludes the possibility of temperature-dependent background. This behavior contrasts with the common observation of a monotonic increase in $f$-spectral weight with decreasing temperature, as previously observed in Ce-based [8–11], Yb-based [54], and U-based [25, 26] HF compounds.

Relocalization of $f$-electron spectral weight has been observed in a limited number of Ce-based HF compounds, such as CePt$_2$In$_7$ [15], CeRu$_4$Sn$_6$ [16], and CeCoGe$_3$ [14]. Several potential origins have been proposed, including the presence of low-energy crystalline-electric-field (CEF) excitations [15], the competition between magnetic order and the Kondo effect [14, 16], and a precursor of the magnetic order [56, 57]. However, in the case of UPd$_2$Al$_3$, the origin of relocalization appears to differ from these Ce-based compounds. First, it is unlikely that low-energy CEF excitations are responsible for relocalization in UPd$_2$Al$_3$, as the relocalized satellite peak is situated away from the $E_F$ [Fig. 4(g)], and the relevant energy scales are different, suggesting a different mechanism. Second, the opening of the hybridization gap at $E_F$ does not seem to lead to relocalization, even though the relocalization temperature coincides with the onset of the hybridization gap [20]. This is because the hybridization process generally promotes an increase in the weight of the $f$-electron spectrum [8].

Since U has three $5f$ electrons, it has been suggested that magnetic order and the Kondo effect involve different $f$ electrons, allowing them to coexist without interference [25, 26, 55]. However, our observations suggest that $f$-electrons with the same orbitals participate in both magnetic ordering and the Kondo effect. This implies that in duality models, the segregation of different U $5f$ orbitals into distinct itinerant and localized components is not necessary.

Li *et al.* proposed that the competition between magnetic order and Kondo effect may lead to relocalization, as they found that the temperature at which the $4f$-spectra weight of CeCoGe$_3$ begins being suppressed coincides with the $T_N$ [14]. In UPd$_2$Al$_3$, consistent with the behavior observed in CePt$_2$In$_7$ [15] and CeRu$_4$Sn$_6$ [16], localization begins at temperatures well above $T_N$. The Knight shift experiment on CePt$_2$In$_7$ [56] and CeRhIn$_5$ [57] demonstrated heavy quasiparticles relocalization above $T_N$, and recent Knight shift mea-surements in UTe$_2$, which lacks magnetic ordering under ambient pressure but exhibits pressure-induced antiferromagnetic order, reveal a similar relocalization of U 5$f$ electrons at low temperatures [58].

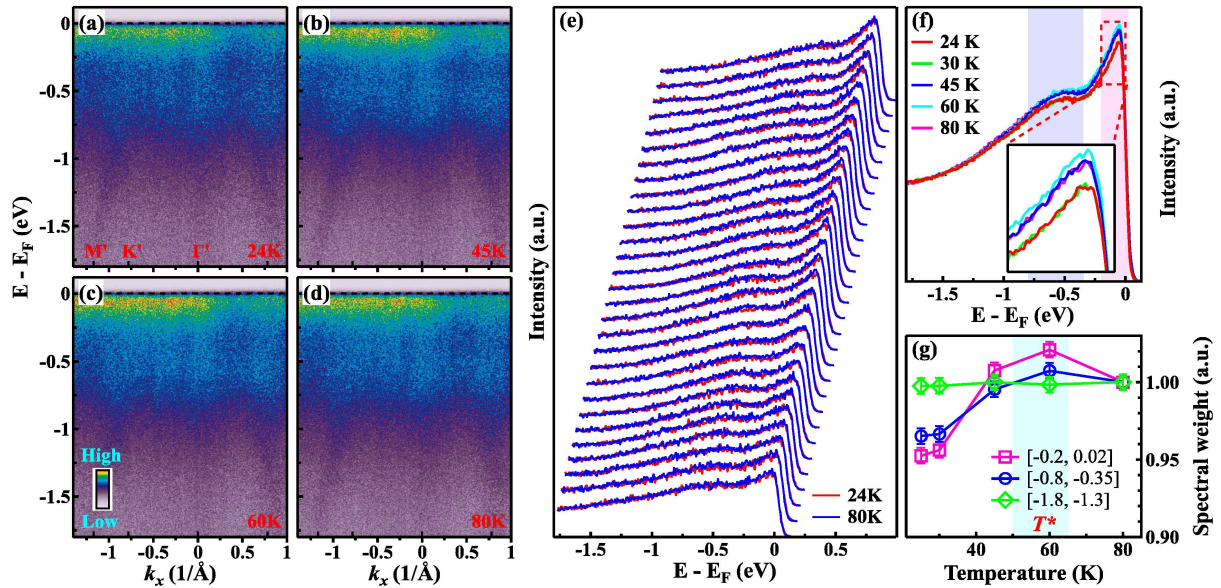

FIG. 4. Temperature dependence of the heavy quasiparticle band of UPd$_2$Al$_3$. (a)-(d) Band structure measured with 108 eV photons at marked temperatures. (e) Detailed ARPES spectra of UPd$_2$Al$_3$ measured at 24 K and 80 K. (f) Angular integrated EDCs at various temperatures. The momentum integrated range is [-1.4 Å$^{-1}$, 1 Å$^{-1}$]. (g) $T$ dependence of the heavy quasiparticle spectral weight. Magenta squares represent itinerant $f$ band at $E_F$ spectral weight integrated over [-200 meV, +20 meV] [light magenta region in (f)], blue circles depict satellite region spectral weight integrated over [-0.8 eV, -0.35 eV] [light blue region in (f)], and green diamonds indicate spectral weight integrals over [-1.8 eV, -1.3 eV]. The momentum integral ranges for three lines are [-0.5 Å$^{-1}$, 0.5 Å$^{-1}$].

Taking into account all the experimental results concerning relocalization mentioned above, which encompass diverse samples and techniques, we propose that the interplay between RKKY interactions and the Kondo effect governs the nuanced temperature-dependent behavior of the $f$ electron spectral weight. Mean-field theory supports the coexistence of RKKY interactions and the Kondo effect well above the $T_N$, given the appropriate Kondo coupling constant ($J$) [3]. As temperature decreases, the Kondo effect leads to the emergence of heavy quasiparticles and the subsequent opening of the hybridization gap, resulting in an increase in the $f$ electron spectral weight. Further cooling, however, magnifies the RKKY interaction, dampening the Kondo effect. If the RKKY interaction becomes sufficiently strong, it can reverse the collective hybridization process, causing the $f$ electrons to relocalize and eventually leading to the establishment of magnetic order at low temperatures. Whether relocalization and/or AFM transition occurs, and the order in which these phenomena are manifested depends on the specific interactions between the RKKY interaction and the Kondo effect. Thus, relocalization of $f$-electrons may also serve as a precursor to the formation of magnetic order [15, 16, 56–58].

## IV. CONCLUSIONS

In conclusion, we investigated the $f$-electron properties of the heavy fermion antiferromagnetic superconductor UPd$_2$Al$_3$. Our measurements reveal that the heavy U 5$f$ bands form at a temperature much higher than the $T^*$. As the temperature decreases, just around $T^*$, the spectrum weight of the 5$f$-band deviates from the behavior of increasing with decreasing temperature and instead weakens with further decreasing. Our work provides clear evidence that (i) the 5$f$ electrons are in an itinerant configuration, and (ii) the relocalization of 5$f$ electrons occurs at low temperatures, resulting from the competition between the RKKY interaction and the Kondo effect, making it a precursor to magnetic order. (iii) The 5$f$ electrons with the same orbital are involved in the Kondo effect and the RKKY interaction, which coexist and compete at low temperatures.

## ACKNOWLEDGMENTS

This work was supported by the National Natural Science Foundation of China (Grant No. 12074436), the Science and Technology Innovation Program of Hunan Province (2022RC3068), and the Natural Science Foundation of Changsha (kq2208254). Work at Los Alamos was performed under the U.S. Department of Energy, Of-fice of Basic Energy Sciences, Quantum Fluctuations in Narrow Band Systems program. J.R. and P.M.O. acknowledge funding from the K. and A. Wallenberg Foundation (Grant No. 2022.0079) and from the Swedish Research Council (VR) (Grant No. 2022-06725).

---

* Corresponding author: jqmeng@csu.edu.cn

[1] Y. Yang, Sci. China-Phys. Mech. Astron. **63**, 117431 (2020).

[2] L. Taillefer, Nat. Phys. **9**, 458 (2013).

[3] S. Doniach, Physica B **91**, 231 (1977).

[4] Y. Haga, E. Yamamoto, Y. Tokiwa, D. Aoki, Y. Inada, R. Settai, T. Maehira, H. Yamagami, H. Harima, and Y. Ōnuki, J. Nucl. Sci. Technol. Suppl. **39**, 56 (2002).

[5] A. Koitzsch, S. V. Borisenko, D. Inosov, J. Geck, V. B. Zabolotnyy, H. Shiozawa, M. Knupfer, J. Fink, B. Büchner, E. D. Bauer, J. L. Sarrao, and R. Follath, Physica C **460-462**, 666 (2007).

[6] K. Götze, J. Klotz, D. Gnida, H. Harima, D. Aoki, A. Demuer, S. Elgazzar, J. Wosnitza, D. Kaczorowski, and I. Sheikin, Phys. Rev. B **92**, 115141 (2015).

[7] K. Knöpfle, A. Mavromaras, L. M. Sandratskii, and J. Kübler, J. Phys.: Condens. Matter **8**, 901 (1996).

[8] Q. Y. Chen, D. F. Xu, X. H. Niu, J. Jiang, R. Peng, H. C. Xu, C. H. P. Wen, Z. F. Ding, K. Huang, L. Shu, Y. J. Zhang, H. Lee, V. N. Strocov, M. Shi, F. Bisti, T. Schmitt, Y. B. Huang, P. Dudin, X. C. Lai, S. Kirchner, H. Q. Yuan, and D. L. Feng, Phys. Rev. B **96**, 045107 (2017).

[9] S. Jang, J. D. Denlinger, J. W. Allen, V. S. Zapf, M. B. Maple, J. N. Kim, B. G. Jang, and J. H. Shim, PNAS **117**, 23467(2020).

[10] Y. Wu, Y. Zhang, F. Du, B. Shen, H. Zheng, Y. Fang, M. Smidman, C. Cao, F. Steglich, H. Q. Yuan, J. D. Denlinger, and Y. Liu, Phys. Rev. Lett. **126**, 216406 (2021).

[11] G. Poelchen, S. Schulz, M. Mende, M. Güttler, A. Generalov, A. V. Fedorov, N. Caroca-Canales, C. Geibel, K. Kliemt, C. Krellner, S. Danzenbächer, D. Y. Usachov, P. Dudin, V. N. Antonov, J. W. Allen, C. Laubschat, K. Kummer, Y. Kucherenko, and D. V. Vyalikh, npj Quant. Mater. **5**, 70 (2020).

[12] S. Kirchner, S. Paschen, Q. Y. Chen, S. Wirth, D. L. Feng, J. D. Thompson, and Q. M. Si, Rev. Mod. Phys. **92**, 011002 (2020).

[13] Y. Z. Zhao, J. J. Song, Q. Y. Wu, H. Liu, C. Zhang, B. Chen, H. Y. Zhang, Z. H. Chen, Y. B. Huang, X. Q. Ye, Y. H. Yuan, Y. X. Duan, J. He, J. Q. Meng, Sci. China-Phys. Mech. Astron. **67**, 247413 (2024).

[14] P. Li, H. Q. Ye, Y. Hu, Y. Fang, Z. G. Xiao, Z. Z. Wu, Z. Y. Shan, R. P. Singh, G. Balakrishnan, D. W. Shen, Y. F. Yang, C. Cao, N. C. Plumb, M. Smidman, M. Shi, J. Kroha, H. Q. Yuan, F. Steglich, and Y. Liu, Phys. Rev. B **107**, L201104 (2023).

[15] Y. Luo, C. Zhang, Q. Y. Wu, F. Y. Wu, J. J. Song, W. Xia, Y. F. Guo, J. Rusz, P. M. Oppeneer, T. Durakiewicz, Y. Z. Zhao, H. Liu, S. X. Zhu, Y. H. Yuan, X. F. Tang, J. He, S. Y. Tan, Y. B. Huang, Z. Sun, Y. Liu, H. Y. Liu, Y. X. Duan, and J. Q. Meng, Phys. Rev. B **101**, 015129 (2020).

[16] F. Y. Wu, Q. Y. Wu, C. Zhang, Y. Luo, X. Q. Liu, Y. F. Xu, D. H. Lu, M. Hashimoto, H. Liu, Y. Z. Zhao, J. J. Song, Y. H. Yuan, H. Y. Liu, J. He, Y. X. Duan, Y. F. Guo, and J. Q. Meng, Frontiers of Physics **18**, 53304 (2023).

[17] Y. P. Liu, Y. J. Zhang, J. J. Dong, H. Lee, Z. X. Wei, W. L. Zhang, C. Y. Chen, H. Q. Yuan, Y. F. Yang, and J. Qi, Phys. Rev. Lett. **124**, 057404 (2020).

[18] Y. H. Pei, Y. J. Zhang, Z. X. Wei, Y. X. Chen, K. Hu, Y. F. Yang, H. Q. Yuan, and J. Qi, Phys. Rev. B **103**, L180409 (2021).

[19] Y. Z. Zhao, Q. Y. Wu, C. Zhang, B. Chen, W. Xia, J. J. Song, Y. H. Yuan, H. Liu, F. Y. Wu, X. Q. Ye, H. Y. Zhang, H. Huang, H. Y. Liu, Y. X. Duan, Y. F. Guo, J. He, J. Q. Meng, Phys. Rev. B **108**, 075115 (2023).

[20] X. D. Yang, P. S. Riseborough, and T. Durakiewicz, J. Phys.: Condens. Matter **23**, 094211 (2011).

[21] S.-i. Fujimori, Y. Saitoh, T. Okane, A. Fujimori, H. Yamagami, Y. Haga, E. Yamamoto, and Y. Ōnuki, Nat. Phys. **3**, 618 (2007).

[22] S.-i. Fujimori, J. Phys. Condens. Mat **28**, 153002 (2016)

[23] S.-i. Fujimori, M. Kobata, Y. Takeda, T. Okane, Y. Saitoh, A. Fujimori, H. Yamagami, Y. Haga, E. Yamamoto, and Y. Ōnuki, Phys. Rev. B **99**, 035109 (2019)

[24] J. Q. Meng, P. M. Oppeneer, J. A. Mydosh, P. S. Riseborough, K. Gofryk, J. J. Joyce, E. D. Bauer, Y. Li, and T. Durakiewicz, Phys. Rev. Lett. **111**, 127002 (2013).

[25] Q. Y. Chen, X. B. Luo, D. H. Xie, M. L. Li, X. Y. Ji, R. Zhou, Y. B. Huang, W. Zhang, W. Feng, Y. Zhang, L. Huang, Q. Q. Hao, Q. Liu, X. G. Zhu, Y. Liu, P. Zhang, X. C. Lai, Q. Si, and S. Y. Tan, Phys. Rev. Lett. **123**, 106402 (2019).

[26] X. Y. Ji, X. B. Luo, Q. Y. Chen, W. Feng, Q. Q. Hao, Q. Liu, Y. Zhang, Y. Liu, X. Y. Wang, S. Y. Tan, and X. C. Lai, Phys. Rev. B **106**, 125120 (2022).

[27] M. K. Liu, R. D. Averitt, T. Durakiewicz, P. H. Tobash, E. D. Bauer, S. A. Trugman, A. J. Taylor, and D. A. Yarotski, Phys. Rev. B **84**, 161101(R) (2011).

[28] G. L. Dakovski, Y. Li, S. M. Gilbertson, G. Rodriguez, A. V. Balatsky, J. Zhu, K. Gofryk, E. D. Bauer, P. H. Tobash, A. Taylor, J. L. Sarrao, P. M. Oppeneer, P. S. Riseborough, J. A. Mydosh, T. Durakiewicz, Phys. Rev. B **84**, 161103(R) (2011).

[29] C. Geibel, C. Schank, S. Thies, H. Kitazawa, C. D. Bredl, A. B Hm, M. Rau, A. Grauel, R. Caspary, R. Helfrich, U. Ahlheim, G. Weber, and F. Steglich, Z. Phys. B Condensed Matter **84**, 1 (1991).

[30] Y. Haga, E. Yamamoto, Y. Inada, D. Aoki, K. Tenya, M. Ikeda, T. Sakakibara, and Y. Ōnuki, J. Phys. Soc. Jpn **65**, 3646 (1996).

[31] M. Jourdan, M. Huth, and H. Adrian, Nature (London) **398**, 47 (1999).

[32] M. Dressel, N. Kasper, K. Petukhov, D. N. Peligrad, B. Gorshunov, M. Jourdan, M. Huth, and H. Adrian, Phys. Rev. B **66**, 035110 (2002).

[33] N. K. Jaggi, O. Mehio, M. Dwyer, L. H. Greene, R. E. Baumbach, P. H. Tobash, E. D. Bauer, J. D. Thompson, and W. K. Park, Phys. Rev. B **95**, 165123 (2017).

[34] Y. Inada, H. Yamagami, Y. Haga, K. Sakurai, Y. Tokiwa, T. Honma, E. Yamamoto, Y. Ōnuki, and T. Yanagisawa, J. Phys. Soc. Jpn **68**, 3643 (1999).

[35] A. Grauel, A. Böhm, H. Fischer, C. Geibel, R. Köhler, R. Modler, C. Schank, F. Steglich, G. Weber, T. Komatsubara, and N. Sato, Phys. Rev. B **46**, 5818 (1992).

[36] A. Böhm, A. Grauel, N. Sato, C. Schank, C. Geibel, T. Komatsubara, G. Weber, F. Steglich, Int. J. Mod. Phys. **7**, 34 (1993).

[37] R. J. Radwanski, R. Michalski, Z. Ropka, Physica B **276-278**, 803 (2000).

[38] R. J. Radwanski, D. M. Nalecz, and Z. Ropka, Acta Physica Polonica A **130**, 545 (2016).

[39] G. Zwicknagl, A. Yaresko, and P. Fulde, Phys. Rev. B **68**, 052508 (2003).

[40] C. Pfleiderer, Rev. Mod. Phys. **81**, 1551 (2009).

[41] J. J. Yeh and I. Lindau, At. Data Nucl. Data Tables **32**, 1 (1985).

[42] J. J. Song, Y. Luo, C. Zhang, Q. Y. Wu, T. Durakiewicz, Y. Sassa, O. Tjernberg, M. Mânsson, M. H. Berntsen, Y. Z. Zhao, H. Liu, S. X. Zhu, Z. T. Liu, F. Y. Wu, S. Y. Liu, E. D. Bauer, J. Rusz, P. M. Oppeneer, Y. H. Yuan, Y. X. Duan, and J. Q. Meng, Chin. Phys. Lett. **38**, 107402 (2021).

[43] T. Takahashi, N. Sato, T. Yokoya, A. Chainani, T. Morimoto, and T. Komatsubara, J. Phys. Soc. Jpn. **65**, 156 (1996).

[44] S.-i. Fujimori, Y. Saito, M. Seki, K. Tamura, M. Mizuta, K. Yamaki, K. Sato, T. Okane, A. Tanaka, N. Sato, T. Komatsubara, Y. Tezuka, S. Shin, S. Suzuki, and S. Sato, J. Electron Spectrosc. Relat. Phenom. **101**, 439 (1999).

[45] E. Guziewicz, T. Durakiewicz, P. M. Oppeneer, J. J. Joyce, J. D. Thompson, C. G. Olson, M. T. Butterfield, A. Wojakowski, D. P. Moore, and A. J. Arko, Phys. Rev. B **73**, 155119 (2006).

[46] G. Zwicknagl, and M. Reese, J. Magn. Magn. Mater. **310**, 201 (2007).

[47] L. Miao, S. Z. Liu, Y. S. Xu, E. C. Kotta, C. J. Kang, S. Ran, J. Paglione, G. Kotliar, N. P. Butch, J. D. Denlinger, and L. A. Wray, Phys. Rev. Lett. **124**, 076401 (2020).

[48] S.-i. Fujimori, Y. Takeda, H. Yamagami, J. Pospíšil, E. Yamamoto, and Y. Haga, Phys. Rev. B **105**, 115128 (2022).

[49] Y. X. Duan, C. Zhang, J. Rusz, P. M. Oppeneer, T. Durakiewicz, Y. Sassa, O. Tjernberg, M. Mânsson, M. H. Berntsen, F. Y. Wu, Y. Z. Zhao, J. J. Song, Q. Y. Wu, Y. Luo, E. D. Bauer, J. D. Thompson, and J. Q. Meng, Phys. Rev. B **100**, 085141 (2019).

[50] Y. H. Yuan, Y. X. Duan, J. Rusz, C. Zhang, J. J. Song, Q. Y. Wu, Y. Sassa, O. Tjernberg, M. Mnsson, M. H. Berntsen, F. Y. Wu, S. Y. Liu, H. Liu, S. X. Zhu, Z. T. Liu, Y. Z. Zhao, P. H. Tobash, E. D. Bauer, J. D. Thompson, P. M. Oppeneer, T. Durakiewicz, and J. Q. Meng, Phys. Rev. B **103**, 125122 (2021).

[51] F. L. Boariu, C. Bareille, H. Schwab, A. Nuber, P. Lejay, T. Durakiewicz, F. Reinert, and A. F. Santander-Syro, Phys. Rev. Lett **110**, 156404 (2013).

[52] See the Supplemental Material for the heavy quasiparticle bands at a higher temperature of 150 K.

[53] C. M. Varma, Rev. Mod. Phys. **48**, 219 (1976)

[54] S. Chatterjee, J. P. Ruf, H. I. Wei, K. D. Finkelstein, D. G. Schlom, and K. M. Shen, Nat. Commun. **8**, 852 (2017).

[55] I. Giannakis, J. Leshen, M. Kavai, S. Ran, C. Kang, S. R. Saha, Y. Zhao, Z. Xu, J. W. Lynn, L. Miao, L. A. Wray, G. Kotliar, N. P. Butch, and P. Aynajian, Sci. Adv. **5**, eaaw9061 (2019).

[56] N. apRoberts-Warren, A. P. Dioguardi, A. C. Shockley, C. H. Lin, J. Crocker, P. Klavins, D. Pines, Y. F. Yang, and N. J. Curro, Phys. Rev. B **83**, 060408(R) (2011).

[57] K. R. Shirer, A. C. Shockley, A. P. Dioguardi, J. Crocker, C. H. Lin, N. apRoberts-Warren, D. M. Nisson, P. Klavins, J. C. Cooley, Y. F. Yang, and N. J. Curro, Proc. Natl. Acad. Sci. USA **109**, E3067 (2012).

[58] N. Azari, M. R. Goeks, M. Yakovlev, M. Abedi, S. R. Dunsiger, S. M. Thomas, J. D. Thompson, P. F. S. Rosa, and J. E. Sonier, Phys. Rev. B **108**, L081103 (2023).

# Supplemental Materials:

## Relocalization of Uranium 5*f* Electrons in Antiferromagnetic Heavy Fermion Superconductor UPd$_2$Al$_3$: Insights from Angle-Resolved Photoemission Spectroscopy

Jiao-Jiao Song,[1] Chen Zhang,[1] Qi-Yi Wu,[1] Yin-Zou Zhao,[1] Ján Rusz,[2] John. J. Joyce,[3] Kevin.S. Graham,[3] Peter Riseborough,[4] Clifford G. Olson,[5] Hao Liu,[1] Bo Chen,[1] Ya-Hua Yuan,[1] Yu-Xia Duan,[1] Eric D. Bauer,[3] Peter M. Oppeneer,[2] Tomasz Durakiewicz,[6] and Jian-Qiao Meng[1,*]

[1]*School of Physics, Central South University, Changsha 410083, Hunan, China*

[2]*Department of Physics and Astronomy, Uppsala University, Box 516, S-75120 Uppsala, Sweden*

[3]*Los Alamos National Laboratory, Los Alamos, New Mexico 87545, USA*

[4]*Temple University, Philadelphia, Pennsylvania 19122, USA*

[5]*Ames Laboratory, Iowa State University, Ames, Iowa 50011, USA*

[6]*Idaho National Laboratory, Idaho Falls, ID 83415 USA*

E-Mail: jqmeng@csu.edu.cn

## Heavy *f*-bands at high temperature

SFigure shows the on-resonance ARPES spectra at 150 K, well above $T^*$. This sample is different from the one used in Figure 4 in the main text. The heavy quasiparticle bands can be clearly seen in SFig. (a) and (b).

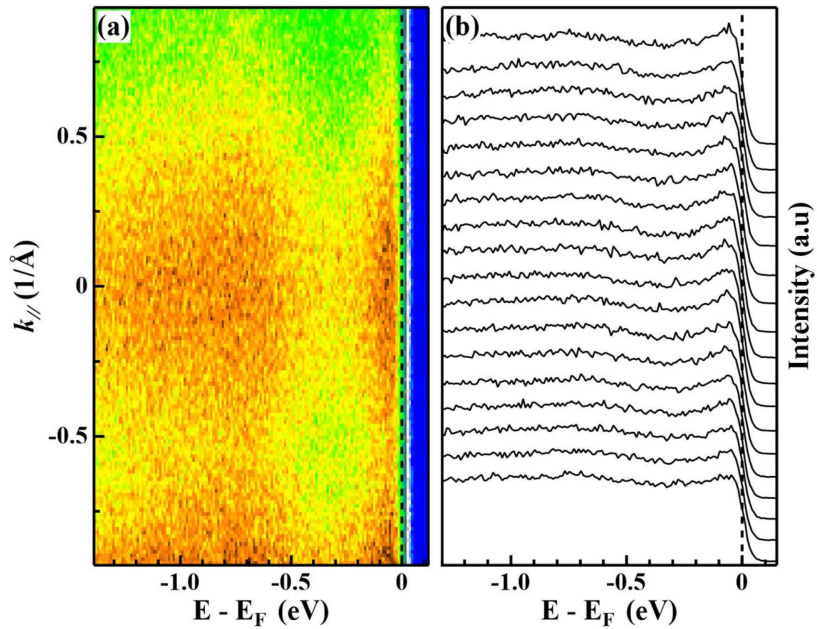

**SFig** (a) on-resonance band structure of UPd$_2$Al$_3$ measured at 150 K. (b) EDCs of the spectra shown in (a).